\documentstyle[11pt,newpasp,twoside]{article}
\newcommand{\be}{\begin{equation}}
\newcommand{\ee}{\end{equation}}
\newcommand{\pprime}{{\prime\prime}}

\reversemarginpar

\begin{document}
\title{Symmetries, magnetic fields, \\ 
and the shape of molecular cloud cores}
\author{Daniele Galli}
\affil{INAF-Osservatorio Astrofisico di Arcetri, Firenze, ITALY}

\begin{abstract}
We develop a formalism wherein the solution of the equations of
equilibrium for a self-gravitating, magnetized interstellar cloud may be
accomplished analytically, under the only hypotheses that ({\it a}\/)
the cloud is scale-free, and ({\it b}\/) is electrically neutral. All
variables are represented as series of scalar and vector spherical
harmonics, and the equilibrium equations are reduced to a set of coupled
algebraic e\-qua\-ti\-ons for the expansion coefficients of the density and
the magnetic vector potential. Previously known axisymmetric solutions
are recovered and new non-axisymmetric solutions are found and discussed
as models for pre-protostellar molecular cloud cores.
\end{abstract}

\section{Antecedents}

{\it ``Do you think that non-symmetric, magnetized, self-gravitating
equilibria can exist?''} This question is a typical example of the
classic, fundamental kind of problems that Frank likes to address. This
contribution presents a summary of the results obtained by the
writer in his effort to answer Frank's question.

\section{Introduction}

Ignoring magnetic fields and considering only bounded masses, the
result that static stars must be spherical probably follows from the
expression of the Newtonian potential alone, but nobody has been able
to give a rigorous proof (even less in general relativity, see Misner,
Thorne, \& Wheeler~1970). For a homogeneous static star, the spherical
configuration is that of minimum gravitational energy (the proof is
due to Lyapunov, and is reported in Poincar\'e~1902). Thus,
the sphere is the only ${\it stable}$ form of equilibrium. Lamb~(1932)
argued that ``the only form of equilibrium of a mass of homogeneous
liquid under its own attraction is a sphere'' on the basis of the
Lichtenstein~(1918) theorem.  As this theorem asserts, a rotating star
is necessarily symmetric with respect to a plane perpendicular to the
rotation axis and passing through the center of mass of the star; if
there is no rotation, every plane through the center of mass must be a
plane of symmetry, hence the star must be spherical.

Interstellar clouds are not homogeneous balls. The most idealized 
model of a cloud consists of an equilibrium configuration of an
isothermal gas, in which self-gravity is balanced by thermal pressure
alone. As it is well known, the equilibrium of such cloud 
is governed by the quasi-linear elliptic partial differential equation (PDE)
\be 
\nabla^2 \psi + e^{\psi}=0,
\label{isot} 
\ee 
where $\psi$ is the non-dimensional gravitational potential.  Does this
equation allow only spherically symmetric solutions? With the condition
$\psi=0$ over an arbitrary boundary surface, eq.~(\ref{isot})
is known to mathematicians as a Liouville-Gel'fand equation. In 1979,
the fundamental theorem of Gidas, Ni, \& Niremberg established that all
solutions of this class of equations are spherically symmetric if the
boundary surface is spherical.


The problem with real clouds is that, unlike stars, they have no
``boundary surface'':  they have infinite extent on the scale of the
forming star.  When the boundary condition on eq.~(\ref{isot}) is pushed
to infinity, non-spherically symmetric solutions of eq.~(\ref{isot})
appear, but they are either unphysical because they correspond to $\rho
<0$ (Matsuno 1987), or not force-free at singular points (Medvedev \&
Narajan 2000). Thus, the deceptively simple question of whether a
balance of gravity and pressure alone may lead to physically acceptable
non-spherical equilibria remains unsettled.
In this contribution, we concentrate on dense, quiescent, molecular
cloud cores, containing a few solar masses of gas, and characterized by
very low levels of turbulence. A wealth of observations shows
that cores are magnetized (see e.g. Crutcher
2001). Therefore, we formulate the problem of the existence of
equilibrium states in the general context of magnetohydrostatics
(MHS).

Little is known about the
intrinsic shapes of cores, which are observed in projection against
the plane of the sky.  From a selected sample, Myers et al. (1991)
concluded that the mean observed aspect ratio was consistent with
either oblate or prolate spheroids. Later, for a larger sample, 
Ryden (1966) and Curry
(2002) found that the observed distribution of axis ratios was more
consistent with prolate rather than oblate shapes.  Recent statistical
studies (Jones, Basu, \& Dubinski 2001; Jones \& Basu 2002), based on
the largest available catalogues of dense cores, 
found unsatisfactory agreement with either oblate or prolate spheroids,
and rejected the possibility that cores are axisymmetric configurations. A
good fit to the observed axial distribution was generally found assuming
that cores are triaxial ellipsoids.  The best-fit axial ratios $a:b:c$
and their standard deviations determined for the same sample by these
studies are remarkably close: $1:(0.9\pm 0.1):(0.5\pm 0.1)$ according
to Jones \& Basu (2002), $1:(0.8\pm 0.2):(0.4\pm 0.2)$ according to
Goodwin, Ward-Thompson, \& Whitworth (2003). The result $a\approx b>c$
clearly suggests that cores are preferentially flattened in one direction
and nearly oblate, and implies that molecular cloud cores may not be
particularly far from conditions of equilibrium.

\section{Ideal magnetohydrostatic equilibria}

Consider a magnetized, isothermal, self-gravitating cloud satisfying
the ideal MHS equations
\be
c_{\rm s}^2\nabla\rho+\rho\nabla {\cal V}=
{1\over 4\pi}(\nabla\times {\bf B})\times {\bf B},
\label{eqforce}
\ee
\be
\nabla\cdot{\bf B}=0,
\label{divB}
\ee
\be
\nabla^2 {\cal V}=4\pi G\rho,
\label{eqPoisson}
\ee
where ${\bf B}$ is the magnetic field, $\rho$ is the density, ${\cal
V}$ is the gravitational potential, and $c_{\rm s}$ is the sound speed.
All known solutions of this set of equations are characterized by a
coordinate symmetry that reduces the number of variables from three to
two or one. The symmetry associated with an ignorable coordinate allows
MHS pro\-ble\-ms to be reduced to a second-order elliptic PDE
(Dungey~1953), conventionally called the Grad-Shafranov 
equation.  Symmetric equilibria may be grouped into ({\it i}\/) {\it
axisymmetric} ($\partial/\partial\varphi=0$), ({\it ii}\/) {\it
cylindrically symmetric} ($\partial/\partial z=0$), and ({\it iii}\/)
{\it helically symmetric} systems
($\partial/\partial\varphi=k\,\partial/\partial z$).  As shown by
Edenstrasser (1980), helical symmetry represents the most general
admissible invariance property of the MHS equations, with rotational
and translational invariance as limiting cases.

Applications to the study of interstellar molecular clouds have 
largeley focused on magnetic configurations of the first kind, 
assuming azimuthal symmetry (e.g. Mouschovias 1976;
Nakano 1979; Tomisaka, Ikeuchi,\& Nakamura~1988).  Cylindrically symmetric
equilibria have been extensively studied, originally in connection with
the stability of galactic spiral arms (Chandrasekhar \& Fermi 1953;
Stod{\'o}{\l}kiewicz 1963), and later as models of filamentary clouds
(Nagasawa~1987; Nakamura, Hanawa, \& Nakano 1995).  In the latter
context, magnetic configurations possessing helical symmetry have 
also been explored (Nakamura, Hanawa, \& Nakano~1993, Fiege \&
Pudritz 2000a,b).  

Even in the special case of non-selfgravitating plasmas, the
fundamental question as to the existence of MHS equilibria of a more
general symmetry or with no symmetry at all (3-D MHS equilibria) has
been formulated several times (e.g. Low 1980, Degtyarev et al. 1985)
but never properly answered. Grad (1967) conjectured that, with ${\cal
V}=0$, only ``highly symmetric'' solutions of the system
(\ref{eqforce})--(\ref{divB}) should be expected, in order to balance
the highly anisotropic Lorentz force with with pressure gradients and
gravity, which are forces involving scalar potentials.  In a
fundamental paper, Parker (1972) proved rigorously the non existence of
3-D MHS equilibria that are small perturbations of 2-D equilibria
having translational symmetry. This result, conventionally referred to
as {\it Parker's theorem}, stimulated much subsequent work and led to
the concept of {\it topological non-equilibrium} for magnetic
configurations lacking high degrees of symmetry (see e.g. Tsinganos,
Distler, \& Rosner 1984; Moffatt 1985, 1986; Vainshtein et al. 2000).
Before extending Parker's theorem and its implications to conditions
typical of molecular cloud cores, one must understand how ({\it a}\/)
fluid motions and ({\it b}\/) gravitational forces affect the existence
of equilibria.

As for the effect of fluid motions, Tsinganos (1982) found that
Parker's theorem remains valid for steady dynamical (${\bf v}\neq 0$)
configurations possessing translational invariance, and stressed the
analogy between this result and the familiar Taylor-Proudman theorem of
hydrodynamics. However, somewhat at variance with these findings, Galli
et al. (2001) found that a class of MHD equilibria possessing axial
symmetry (rotating, magnetized, self-gravitating singular isothermal
disks) does have neighboring non-axisymmetric equilibrium states,
provided the rotation speed becomes sufficiently high
(supermagnetosonic). As for the effect of gravity, Field (1982, quoted
by Tsinganos et al. 1984) noted that the neglect of the constraining
effect exerted by the plasma's self-gravity may severely limit the
existence of MHS equilibria.  In a series of papers (Low 1985; Bogdan
\& Low 1986; Low 1991), Low and collaborators have elaborated a general
method to solve the equations of MHS in the presence of an external
gravitational field, either uniform or variying as $1/r^2$, without
postulating the existence of an ignorable coordinate but assuming an
electric current everywhere perpendicular to the gravitational field.
Even under these restricted assumptions, the problem presents
however considerable mathematical difficulties.

\section{Scale-free equilibria}

In order to simplify the problem, we assume that the density and
magnetic field are scale-free, i.e. varying as powers of the radial
coordinate $r$ in spherical coordinates $(r,\theta,\varphi)$. 
In general, any solenoidal vector field {\bf B} can be expressed as a
linear combination of two basic fields ${\bf T}$ and ${\bf S}$
(Chandrasekhar 1961), characterized by vanishing radial
component ($T_r=0$) and vanishing radial component of the curl
($[\nabla\times {\bf S}]_r=0$), respectively.  For a scale-free
configuration, if there is no current at one radius, then the same is
true for all radii. This condition must be certainly satisfied at large
radii, otherwise there would be a flow of charge to infinity, and the
cloud would become electrically charged. Thus, the magnetic field in
our problem must be purely of the ${\bf S}$-type, and given by
\be
{\bf B}=\nabla\times\left[\nabla\left({\Psi\over r}\right)
\times {\bf r}\right],
\label{defB}
\ee
where $\Psi(r,\theta,\varphi)$ is an arbitrary scalar function of position.
Here we adopt the scaling of Li \& Shu~(1996, hereafter LS96),
\be
\rho={c_{\rm s}^2\over 2\pi Gr^2} R(\theta,\varphi),
\label{scalrho}
\ee
\be
{\cal V}=2c_{\rm s}^2[(1+H_0)\ln r+V(\theta,\varphi)],
\label{calV}
\ee
\be
\Psi={2 c_{\rm s}^2r\over \sqrt{G}}F(\theta,\varphi).
\label{scalPsi}
\ee
The quantity $H_0$ in the expression of the gravitational potential
eq.~(\ref{calV}) is a dimensionless constant measuring the fractional
increase in the mean density that arises because the magnetic field
contributes to support the cloud against self-gravity (LS96).

\section{Spectral magnetohydrostatics}

A natural basis for a multipole expansion of
eq.~(\ref{eqforce})--(\ref{eqPoisson}) is provided by {\it scalar
spherical harmonics} $Y_{lm}$ and 
{\it vector spherical harmonics}
${\bf P}_{lm}$, ${\bf B}_{lm}$, 
and ${\bf C}_{lm}$ (see e. g. Morse \& Feshbach 1953)
\be
{\bf P}_{lm}=Y_{lm}(\theta,\varphi){\bf\hat r},~~~~{\bf B}_{lm}=
{1\over\sqrt{l(l+1)}}\nabla_\Omega Y_{lm}(\theta,\varphi),~~~~{\bf C}_{lm}=
-{\bf\hat r}\times {\bf B}_{lm},
\label{vectharm}
\ee
where $\nabla_\Omega$ is the angular part of the $\nabla$ operator.
We expand the functions $R(\theta,\varphi)$, 
$V(\theta,\varphi)$, and 
$F(\theta,\varphi)$ in scalar spherical harmonics,
\be
R(\theta,\varphi)=1+H_0+\sum_{lm}R_{lm}Y_{lm}(\theta,\varphi),
\label{Rexp}
\ee
\be
V(\theta,\varphi)=\sum_{lm}V_{lm}Y_{lm}(\theta,\varphi),\qquad
F(\theta,\varphi)=\sum_{lm}F_{lm}Y_{lm}(\theta,\varphi),
\label{Fexp}
\ee
where the sum is for $l\ge 1$ and $-l\le m \le l$, and 
$R_{lm}$, $V_{lm}$, and $F_{lm}$ are constant complex coefficients.
This expansion makes possible to solve
immediately Poisson's equation, obtaining
$R_{lm}=-l(l+1)V_{lm}$.
Vector quantities like $\nabla_\Omega R$, $\nabla_\Omega V$, etc.,
are instead expressed as series of vector spherical harmonics, as
\be
\nabla_\Omega R=\sum_{lm}\sqrt{l(l+1)}R_{lm}{\bf B}_{lm},\quad
\nabla_\Omega V=\sum_{lm}\sqrt{l(l+1)}V_{lm}{\bf B}_{lm},~\rm{etc.}
\ee
The properties of the product of vector spherical
harmonics allow non-linear terms to be dealt with in a systematic way
in terms of Wigner $3j$ symbols (generalized Clebsh-Gordan
coefficients), that can be easily evaluated by Racah's formula
(see e.g. Varshalovich, Moskalev, \& Khersonskii 1988).

After several manipulations (the details will be given elsewhere), 
we obtain the equation expressing the condition of radial force balance,
\be
H_0 R_{lm} = \sum_{l^\prime m^\prime}\sum_{l^\pprime m^\pprime}
\Theta^{m m^\prime m^\pprime}_{l l^\prime l^\pprime}
A_{l^\prime m^\prime}A_{l^\pprime m^\pprime},
\ee
the equation expressing the condition of tangential force balance,
\be
[l(l+1)-2(1+H_0)] R_{lm} = \sum_{l^\prime m^\prime}
\sum_{l^\pprime m^\pprime} 
(\Delta^{m m^\prime m^\pprime}_{l l^\prime l^\pprime}R_{l^\prime m^\prime}R_{l^\pprime m^\pprime}
+\Gamma^{m m^\prime m^\pprime}_{l l^\prime l^\pprime}A_{l^\prime m^\prime}A_{l^\pprime m^\pprime})
\ee
and the condition for magnetic support,
\be
\sum_{lm} |A_{lm}|^2=4\pi H_0(1+H_0),
\ee
where $R_{lm}$ and $A_{lm}=l(l+1)F_{lm}$ are complex
coefficients.  The coupling coefficients $\Theta^{m m^\prime
m^\pprime}_{l l^\prime l^\pprime}$, $\Delta^{m m^\prime m^\pprime}_{l
l^\prime l^\pprime}$, and $\Gamma^{m m^\prime
m^\pprime}_{l l^\prime l^\pprime}$ are real, and
are equal to zero when the usual triangular conditions on the Wigner
$3j$ symbols are not all satisfied.  In this way, the representation of the
physical variables has been transferred from function space, in terms of
$\theta$ and $\varphi$, to the infinite-dimensional Hilbert space, each
vector component now being the coefficient of the corresponding
harmonic.


\section{Test of the method: the axisymmetric case}

The procedure for examining the properties of the spectral equations
is to truncate the equations by selecting a finite set of expansion
coefficients up to some $l=l_{\rm max}$, and setting to zero
all other coefficients.  As a test of the method, we have solved the
spectral MHS equations for $m=0$ and various values of $l_{\rm max}$,
and compared the results with the axisymmetric solution (singular
isothermal toroids) found by LS96. Setting $l_{\rm max}=2$,
is equivalent to solving only the radial component of the equation of
force balance. At this level of approximation, all variables 
have simple analytical expressions.
For increasingly larger values of $l_{\rm max}$, we solve numerically
the resulting nonlinear algebraic system with
Newton's method, assuming as a first guess the solution obtained with
the previous value of $l_{\rm max}$.  Already for $l_{\rm max}=4$, the
solution is within $1\%$ of the exact solution obtained with the method
of LS96 over a wide range of values of $H_0$.

\subsection{Perturbation analysis}

In order to assess the validity of Parker's theorem for models of
molecular cloud cores, we investigate whether the class of axisymmetric toroids
of LS96 can allow neighbouring 3-D equibria. To this goal, we perform
a linearization of the set of nonlinear algebraic equations near the
solution determined in the previous section.  We consider two kinds of
perturbations: ({\it i}\/) azimuthally symmetric but
non-symmetric with respect to the $\theta=\pi/2$ plane, and ({\it ii}\/)
symmetric with respect to the $\theta=\pi/2$ plane, but
not azimuthally symmetric. In the former case the density perturbation is
characterized by $l$ odd and $m=0$, in the latter by $l+m$ even and $m\neq
0$. The details of the calculation are too lenghty to be given here,
and will be published elsewhere. 

The results show that the axisymmetric
singular isothermal toroids {\it allow} neighbouring 3-D equilibria of
either type ({\it i}\/) or ({\it ii}\/), for {\it any} value of $H_0$,
if the perturbation of the density contains harmonics with $l=1$ (for case
{\it i}\/) or harmonics with $m=1$ (for case {\it ii}\/). The first case
corresponds to a ``bending'' of the toroid with respect to the original
equatorial plane, the second case corresponds to a ``stretching'' of
the toroid on one side, while preserving equatorial symmetry.
In the limiting case $H_0=0$, where the toroid becomes the unmagnetized
singular isothermal sphere, these two distortions are equivalent, and
correspond to the small ellipticity limit of the ellipsoidal equilibria
found by Medvedev \& Narajan~(2000).  In the limit $H_0\rightarrow\infty$,
the toroid becomes a thin disk, and the $m=1$ distortion reproduces
the small ellipticity limit of the sequence of elliptical disklike
equilibria found by Galli et al.~(2001).  Since the ellipsoidal
equilibria by Medvedev \& Narajan~(2000) are not force-free at the
central singularity, the same is probably true for the elliptic
disklike equilibria found by Galli et al.~(2001) and all intermediate
toroidal configurations with a $m=1$ distortion.  Very likely, also
the ``bending'' distortion of the toroids results in an equilibrium
configuration that is not force-free at the central singularity, but
a rigorous proof would be beyond the goals of this paper.  We provisionally
dismiss both the ``$l=1$ bending'' and ``$m=1$ stretching'',
although the appearance of these permitted distortions of a singular
isothermal system deserves further scrutiny.

For arbitrary $m\neq 1$, the perturbation analysis shows that the
singular isothermal toroids of LS96 do {\it not} have neighbouring
non-axisymmetric equilibria. This results is in agreement with Parker's
theorem, and extends its validity for the first time to the realm of
self-gravitating equilibria. Combining this result with the findings of
Galli et al.~(2001), one is led to the conclusion that the presence of
fluid motions (specifically, rotation), rather than the effect of
self-gravity, is a crucial ingredient for the occurrence of symmetry
breakings in MHD equilibria.  However, the required supermagnetosonic
levels of rotation are too high to be of any consequence for molecular
cloud cores.

What is then the fate of a quasi-static cloud core and its frozen-in
magnetic field, formed by whatever process in a configuration presumably
lacking a high degree of symmetry? Parker (1979) argued that realistic
magnetic fields with no well defined symmetries must evolve in a
genuinely time-dependent way, until all non-symmetric components of the
field are destroyed by dissipation and reconnection and the topology
becomes symmetric.  The numerical simulations of Vainshtein et al.~(2000)
provide a striking illustration of this process.

The problem with starless cores is that ohmic dissipation times are of the
order of $\sim 10^{15}$~yr, and therefore the kind of monotonic relaxation
to equilibrium envisaged by Parker can be ruled out.  To the extent
that cores can be represented as isolated systems devoid of internal
sources of energy, these systems must instead undergo weakly damped Alfv\`en
oscillations around the equilibrium state (say, a non-singular version
of the axisymmetric toroids of LS96) with period $\sim L/(c_{\rm s}^2
+v_{\rm A}^2)^{1/2}$ (Woltjer~1962), where $v_{\rm A}$ is the Alfv\`en speed of
the plasma.  As we noted in \S~2, the deviation of the {\it average} shape
of cloud cores from axisymmetric oblate spheroids is small ($a\approx
b$). If we adopt this deviation as a rough measure of nonequilibrium,
a simple harmonic oscillator analogy provides an estimate of the average
velocity $\langle v\rangle$ of pulsation,
\be
\frac{\langle v\rangle}{(c_{\rm s}^2+v_{\rm A}^2)^{1/2}}
\approx 
\frac{\langle a-b\rangle}{a},
\ee
which, for  $a:b=1:0.8$ (see \S~2) gives $\langle v\rangle\approx
0.2\;({c_{\rm s}^2+v_A^2})^{1/2}$. For a typical core with $c_{\rm s}=0.2$~km~s$^{-1}$,
$n({\rm H}_2)=10^4$~cm$^{-3}$, and $B=10$~$\mu$G, this implies coherent
pulsations with $\langle v\rangle\approx 0.05$~km~s$^{-1}$. 

\section{Conclusions}

Whether or not real cloud cores are triaxial ellipsoids that pulsate
because of their intrinsic topological non-equilibrium, as suggested by
the theoretical arguments discussed in the previous
sections, the material presented in this contribution illustrates well the
richness of possible implications concealed in a typical question asked
by Frank.

\end{document}